\def\kms{\mbox{\,km\,s$^{-1}$}}

\documentstyle[11pt,paspconf,epsf]{article}

\begin{document}

\title{Mass-Models of Five Nearby Dwarf Irregular Galaxies}
\author{St\'ephanie C{\^o}t\'e}
\affil{European Southern Observatory, Karl-Schwarzschild-Str. 2,
D-85748, Garching bei M\"unchen, Germany}
\author{Ken Freeman}
\affil{Mount Stromlo Observatory, Weston Creek ACT 2611, Australia}
\author{Claude Carignan}
\affil{Universit\'e de Montr\'eal, CP 6128, succ. A, Montr\'eal, H3C
3J7, Canada}

\begin{abstract}

Five nearby dwarf irregular galaxies, amongst the recently surveyed
dwarf members of the Sculptor and Centaurus A groups (at 2.5 Mpc and 3.5
Mpc), have been imaged in neutral hydrogen (HI) with the Australia
Telescope and the Very Large Array. These objects have absolute magnitudes
$M_B$ in the range -14.9 to -12.7, yet they are clearly rotationally 
supported, with maximum rotation velocities ranging from 43 \kms to 67
\kms . Multi-component mass-models have been fitted to the rotation
curves. We investigate the properties of their dark matter halos, and
the scaling laws of the dark matter halos parameters.

\end{abstract}

\section{Introduction}

The strongest evidence for dark matter in galaxies comes from extended
neutral hydrogen (HI) rotation curves of galaxies, and especially
amongst all the galaxy types from dwarf Irregulars (dIrrs) rotation curves.
These systems are literally dominated by dark matter, their luminous
matter usually bring only a minor dynamical contribution. From their extended
HI distribution one can derive rotation curves to large galactocentric
radii, probing very far out into the dark halo potential. For these
reasons the dark matter halo parameters can be tightly constrained.
 And by studying extreme low-mass dIrrs one gets a better handle on the
 dark halo scaling laws, since there are known correlations of the
 dark halo properties with galaxy types (Kormendy 1987).

The  dIrrs of our sample are dwarf members of
the two nearest groups of galaxies outside the Local Group, Sculptor at 2.5 Mpc
and Centaurus A at 3.5 Mpc. Our Parkes HI survey detected three dozens
of dwarf members in these two groups (see C\^ot\'e et al 1996 for a
detailed description). Five objects amongst the most gas-rich ones, 
 with absolute magnitude M$_B$
in the range $-15$ to $-12.7$, were selected for kinematical studies:
 UGCA 442 (Sculptor Group), ESO381-G20, DDO 161, ESO444-G84, and ESO325-G11 
 (Centaurus A Group). Some of their optical parameters are listed
 in Table 1.

\begin{table}
\caption{Optical parameters of the selected dwarfs}
\begin{center}
\begin{tabular}{lccccc}
Name  & R.A. \& Dec & M$_B$ & B-R & $\mu _B$ & $\alpha ^{-1}$\\
      &  (1950)  &     &       & mag/arcsec$^2$ & kpc \\
\tableline
UGCA 442 & 23 41 10 -32 13 58 & -13.8 & 0.58 & 22.2 & 0.43 \\
ESO381-G20 & 12 43 18 -33 33 54 & -13.9 & 0.59 & 22.9 & 0.62 \\
DDO 161 & 13 00 38 -17 09 14 & -14.9 & 0.71 & 21.8 & 0.7 \\
ESO444-G84 & 13 34 32 -27 47 30 & -12.7 & 0.28 & 23.1 & 0.24 \\
ESO325-G11 & 13 42 01 -41 36 30 & -13.8 & 0.69 & 24.0 & 1.2 \\
\end{tabular}
\end{center}
\end{table}

\section{HI observations}

HI mappings were done with the Australia Telescope (1.5km \& 3km arrays)
and the Very Large Array (B/C \& C/D), providing velocity resolutions of
3.3 \kms and 5.2 \kms respectively, and beam sizes ranging from 13" to
40". The HI 
distributions extend well beyond the optical galaxies in all cases, out
to 2 Holmberg radii on average, which means for our dwarfs radii between
4 and 13 $\alpha ^{-1}$ (where $\alpha ^{-1}$ is the optical disk
scalelength). 
Their velocity fields show the clear signature of large-scale rotation. 
These dwarfs are therefore gravitationally supported by rotation rather
than pressure-supported by random motions (since their velocity
dispersions are much lower, see below). For lower luminosity systems
the maximum rotation velocity ($V_{max}$) decreases so one expects that
eventually the random motions dominate in very small galaxies. Here the
range of magnitudes of our objects overlap with the Lo et al 1993 Local
Group sample which they found to be pressure-supported, although 
Hoffman et al 1996 recently got higher $V_{max}$ for half of these dwarfs from
Arecibo mapping. So it seems that down to at least $M_B=-12$ the systems
are supported by rotation. 

Inclined tilted-rings were fitted to the velocity fields to yield
the rotation curves, allowing to model the warps which are found to be
about 10\deg on average (in position angle and inclination), leaving
low velocity residuals of the order of 5 \kms . As is expected for such
systems the rotation curves are seen to be slowly rising (see Figure 1),
however the flat part is reached in all cases, and the 
 V$_{max}$ range from 43 \kms ~to 67 \kms .
The velocity dispersions are mostly uniform with an average value of 
8 \kms, which is similar to giant spirals where $\sigma \sim $ 10 \kms 
 (Shostak \& van der Kruit 1984). So indeed their $V_{max}$ is several
 times higher than their $\sigma $ so that they are supported mainly
 by rotation, which  
  allows us to build valid mass-models with
 their rotation curves, provided the velocities are corrected for
 asymmetric drift to take this pressure term into account (Skillman
 1987).

\begin{figure}
\plottwo{stefcen0.ps}{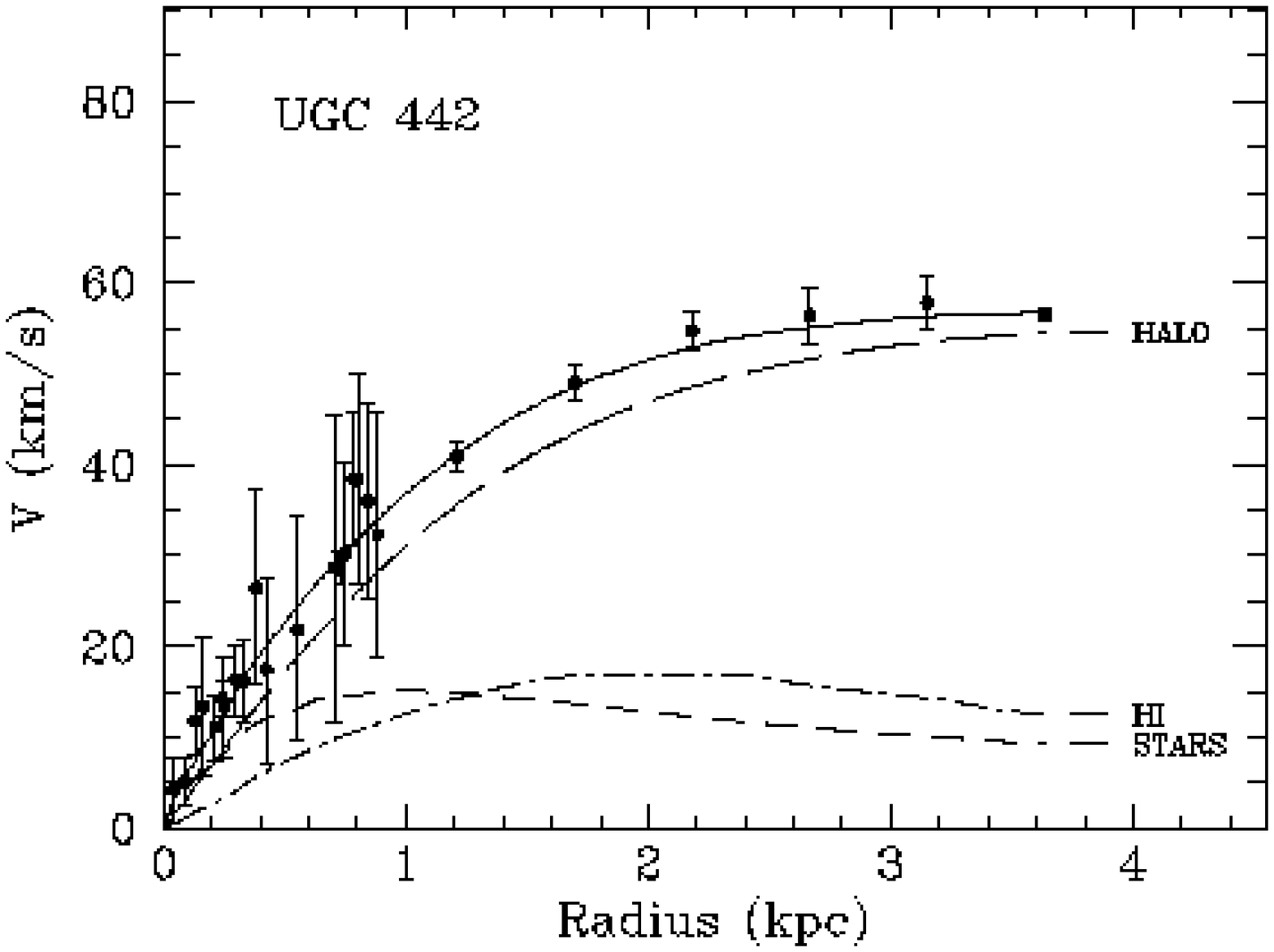}
\plottwo{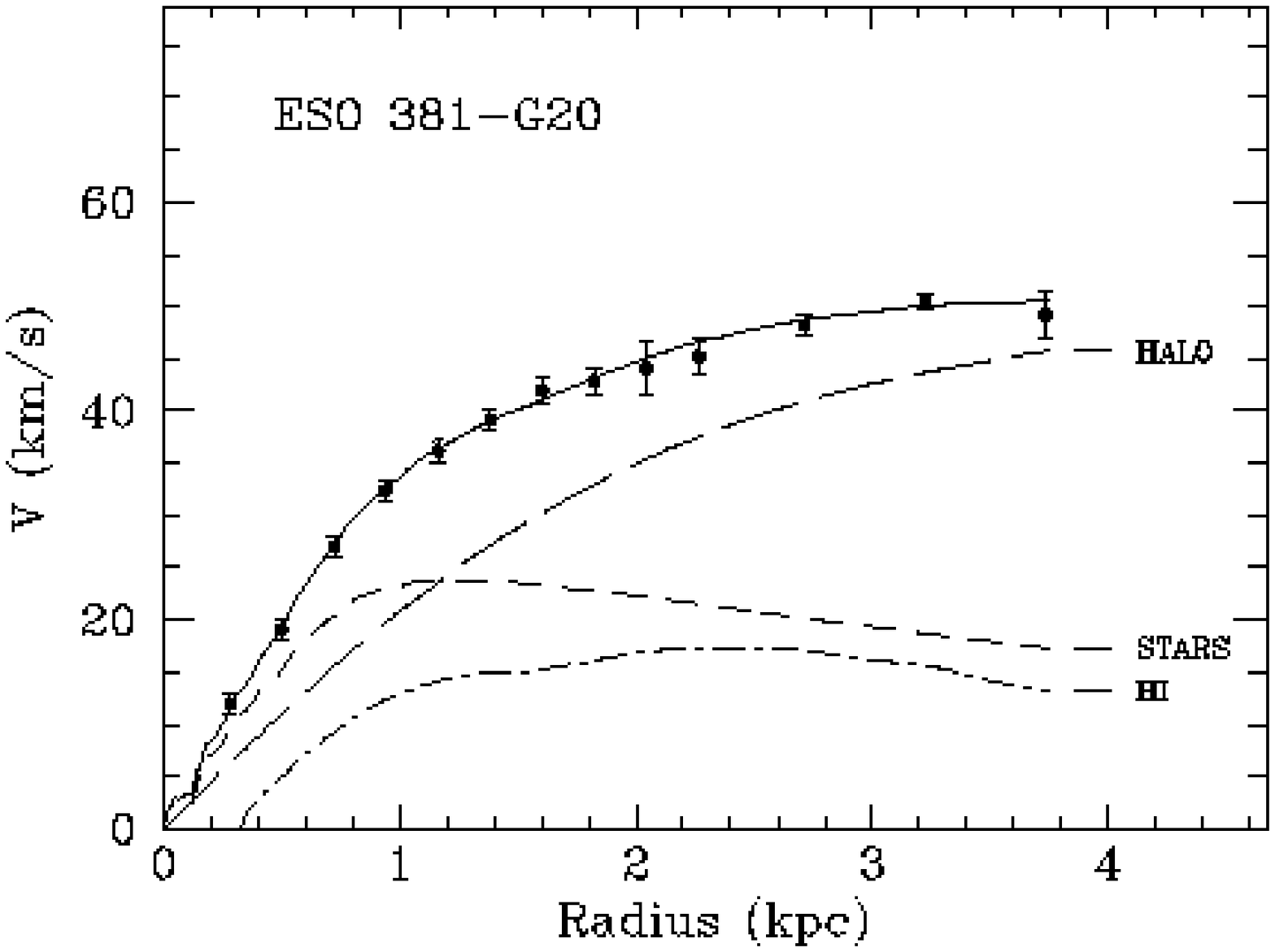}{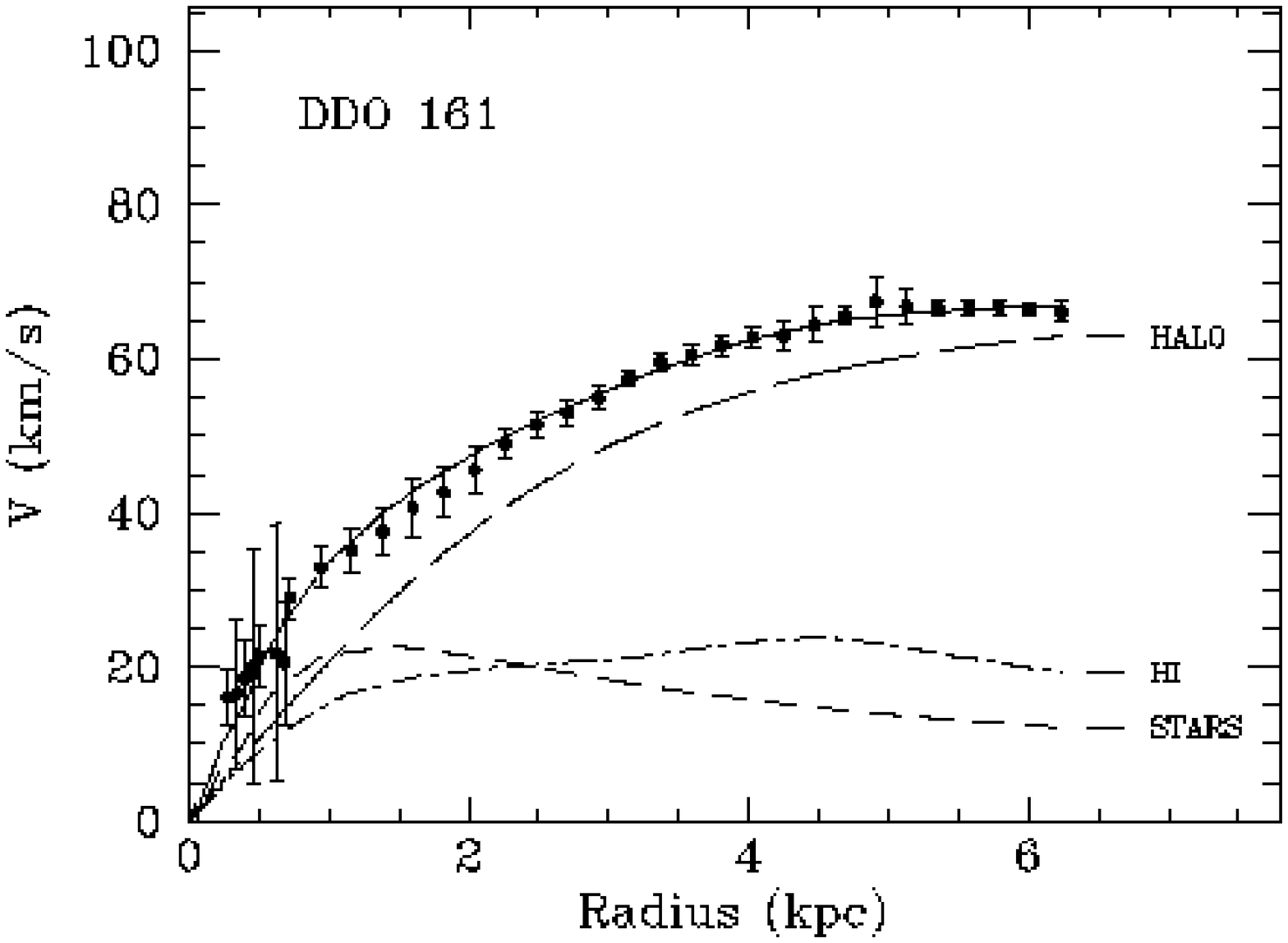}
\plottwo{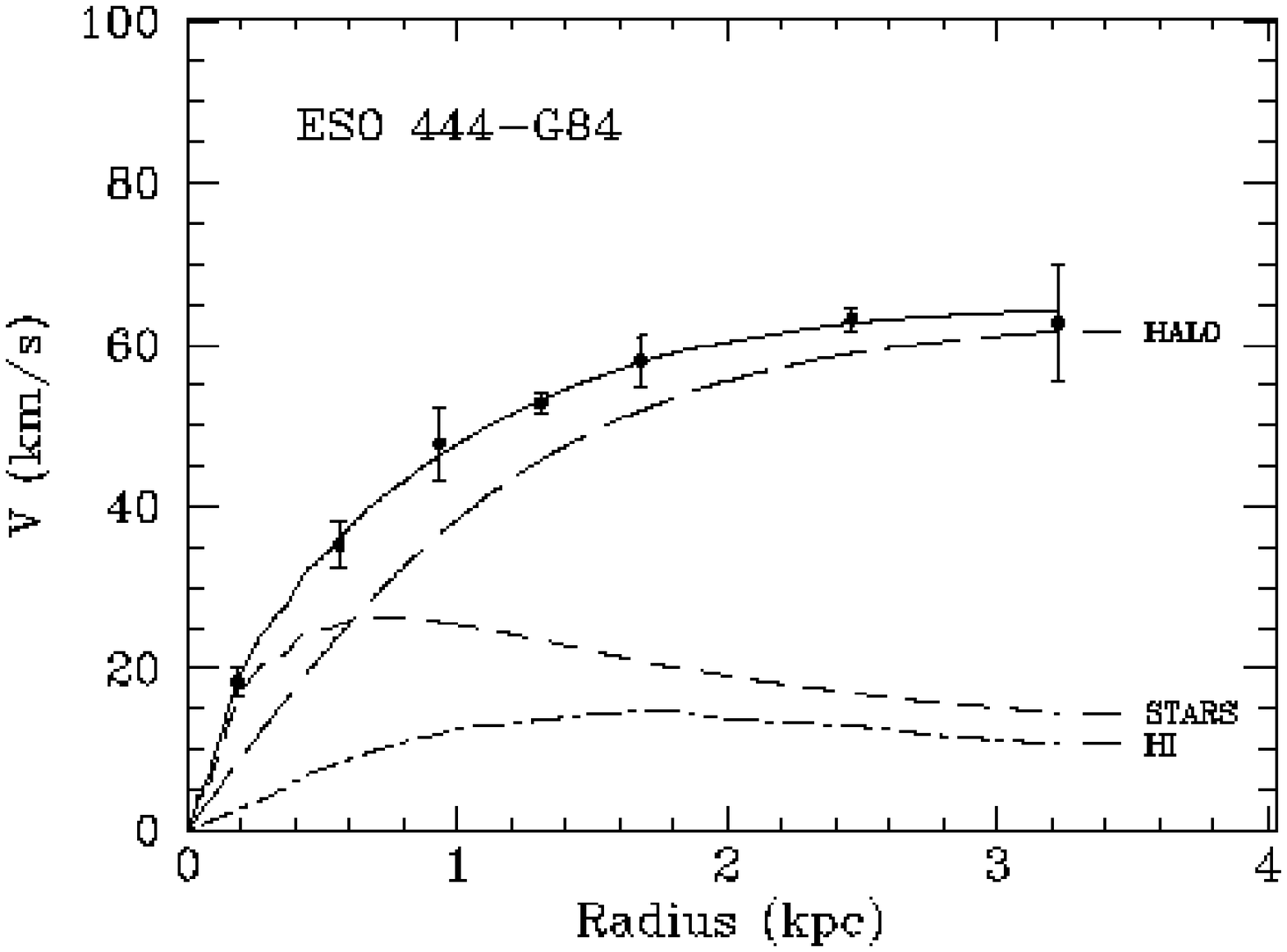}{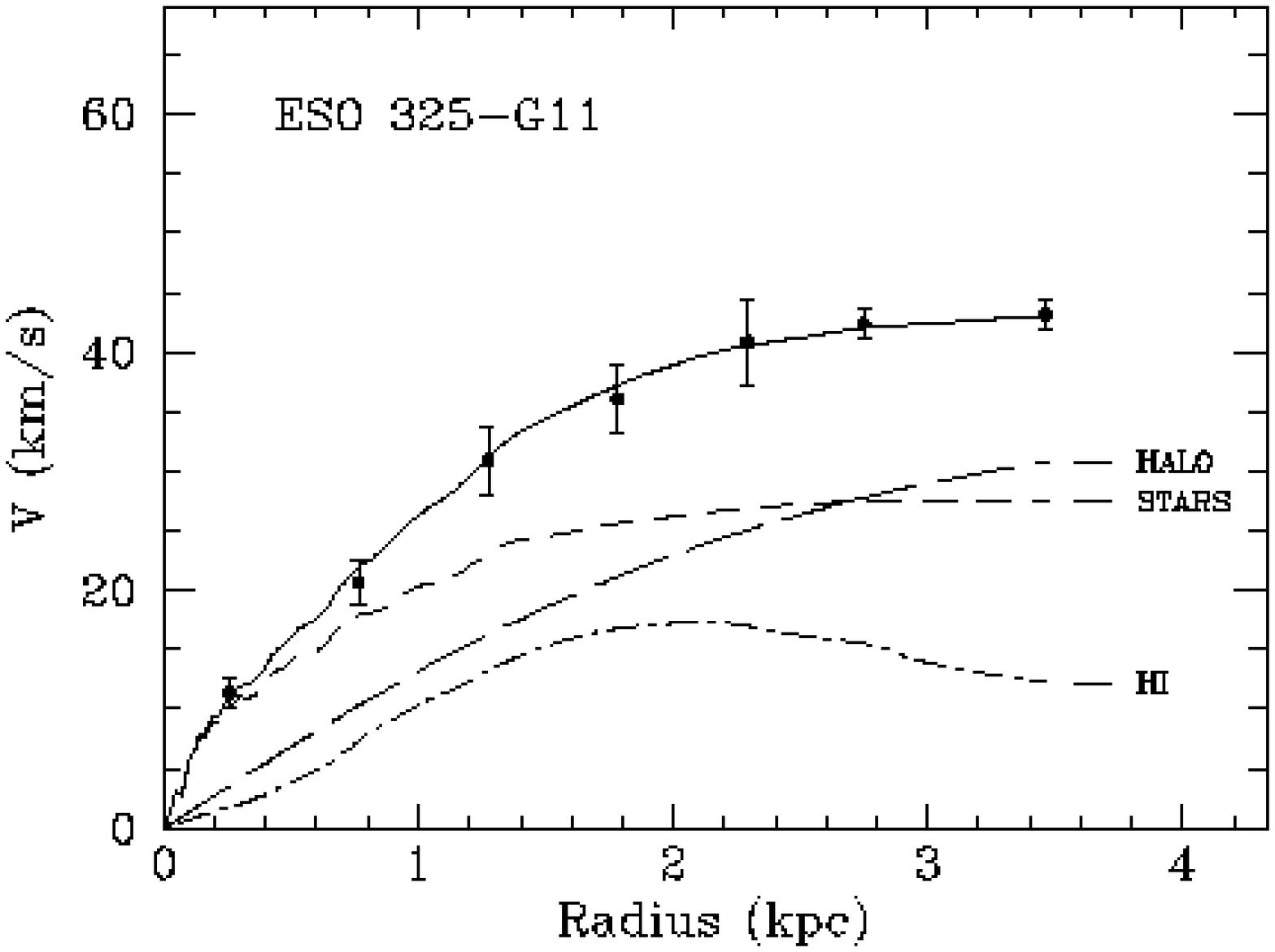}
\caption{Mass-models for the five dIrrs, showing the contributions from
the stellar disk, the HI disk and the dark halo in each model.}
\end{figure}

\section{Mass-Models}

Multi-component mass-models have been fitted to the rotation curves
(Figure 1),
which assume the mass distribution of a galaxy to consist of 
 a stellar disk, a neutral gas disk, and a dark matter
halo. The mass-components of the luminous material (stars and gas) are
calculated from the surface brightness profile and the HI radial
surface density profile, while a nonsingular isothermal sphere is used
for the dark matter halo (see Carignan 1985).
 The mass-to-light ratio
of the stellar disk $(M/L)_{lum}$ is a free parameter, it is applied to
the luminosity profiles (here obtained in I with the Siding Spring 2.3m
telescope)
to obtain the stellar mass surface densities. The dark matter halo has
two free parameters, its core radius and its velocity dispersion, with
its central density given by 
$\rho _0 = {{9 \sigma ^2}/{4 \pi G r_c^2}}$. Of course the more 
extended is the rotation curve the better one can hope to constrain
these parameters. Also combining the HI rotation curves with optical 
H$\alpha $ rotation curves in the inner parts,
like we have here for UGCA 442 and DDO 161, helps better constraining 
$(M/L)_{lum}$. 

For each dwarf a `best-fit' model and a so-called 
`maximum-disk' model ($(M/L)_{lum}$ being pushed to its maximal allowed
value) were constructed. Figure 1 shows the best-fit models. It is clear
 that the dark matter is in every
case the most massive component, accounting for
at least 54\% and up to more than 92\% of their total mass (out to the
last measured point of their rotation curves).
 They are definitely dark-halo dominated, and in fact even in
the rising part of the rotation curve, except for ESO325-G11, the dark
halo becomes already the major dynamical contributor and the stellar
disk is not self-gravitating. This is also the
case for other dwarfs like DDO 154 (Carignan \& Freeman 1988), DDO 127
(Kormendy 1996) and DDO 170 (Lake et al 1990) in which the rotation
curve is explained by dark matter even near the center.
Even the gas component is sometimes more dynamically
significant than the stellar disk, like in DDO 161 or UGCA 442 which has
two times more mass in HI than in stars. So even if $(M/L)_{lum}$ is the
least well constrained parameter in our models, this does not impact very 
much on the total mass or mass-to-light ratio $(M/L_B)_{dyn}$ derived for
these objects.

\section{Properties of dark halos}

Let us now compare the dark halo parameters of our dwarfs with those of
normal spirals, in order to inspect the halo scaling laws. Kormendy
(1990) pointed out that the central halo density seems to increase for
galaxies of decreasing absolute magnitude. With their study of DDO 154
Carignan \& Freeman (1988) suggested that the ratio of dark to luminous
matter ($M_{dark}/M_{lum}$) gets larger for galaxies at the low mass
end.
Our mass-models results confirm that the total
mass-to-light ratio scales with luminosity, in the sense of course that lower
luminosity galaxies have a higher ratio of dark matter mass to luminous
mass than so-called normal galaxies. This is true even when comparing 
their dark matter and luminous masses at a particular radius, for
example at a few times the stellar disk scalelength $\alpha ^{-1}$
rather than at the last measured point at r$_{max}$ 
(otherwise galaxies with more extended HI rotation curves will be
advantaged as more of their dark halo is sampled). In Figure 2 
$M_{dark}/M_{lum}$ at a radius of 7 $\alpha ^{-1}$ is plotted for
a sample of galaxies for which rotation curves extend at least that far
(from the compilation of Broeils 1992). The results of our dwarfs
maximum-disk models are added (with lower limits in two cases because
their rotation curves don't reach 7 $\alpha ^{-1}$). 

It can be argued
that comparing these masses at a certain optical radius, like $R_{25}$
or 7 $\alpha ^{-1}$ here, is perhaps not the best choice considering
that the stellar disk is so unimportant for dwarfs and therefore these
values are not representative of the baryonic scalelength for a dwarf.
Nevertheless despite this the trend is clearly visible in Figure 2. For
normal spirals $M_{dark}/M_{lum}$ is not far from  $\sim $1 (which was
noticed many years ago by Bahcall \& Casertano 1985 for example), but is
known to be a function of luminosity ({\it eg.} Persic \& Salucci 1988,
1990).
At lower luminosity this increases very rapidly. The point at the top is
DDO 170 (Lake et al 1990); also NGC 5585 (C\^ot\'e et al 1991) has a
high $M_{dark}/M_{lum}$ of 6.9, as most low-surface-brightness galaxies
seem to have higher fraction of dark matter (see also de Blok, this
volume). One also notices that DDO 154 (at M$_B$=-13.8) is not an
exceptional galaxy, other dwarfs of the same luminosity range have
similar (or even more extreme!) dark matter properties.

%\begin{figure}
%\plotone{fig2.ps}
%\end{figure}

\begin{figure}[h]
\epsfxsize=8.5cm
\epsfysize=7cm
\centerline{\epsffile{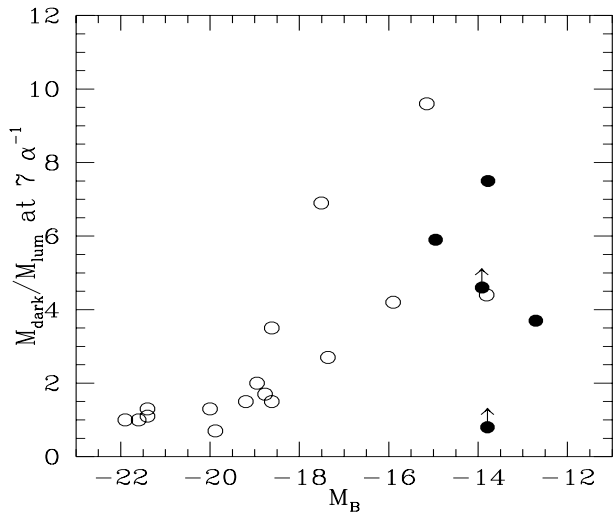}}
\caption[1]{\label{f2} 
Ratios of dark to luminous mass at a radius of 7 $\alpha
^{-1}$. Filled circles are for our sample, open circles for the galaxies
compiled by Broeils 1992.}
\end{figure}

\begin{figure}[h]
\epsfxsize=8.5cm
\epsfysize=7cm
\centerline{\epsffile{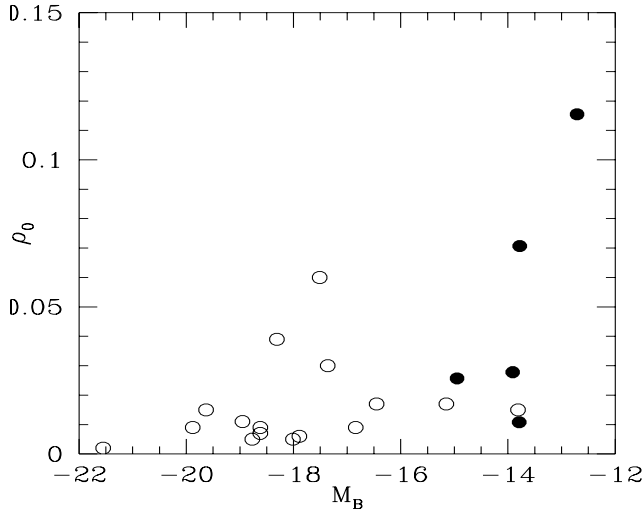}}
\caption[1]{\label{f3} 
Central dark halo densities (in M$_\odot$ pc$^{-3}$) for our sample 
(filled circles) and
for similarly modeled galaxies, compiled in C\^ot\'e 1995 (open
circles).}
\end{figure}

 Looking now at the dark matter halo
parameters, in Figure 3 the central dark halo density is plotted for our
dwarfs as well as a sample of 16 galaxies which have been modeled
similarly using an isothermal sphere for the dark halo (mainly from the
Puche \& Carignan 1991 compilation, see C\^ot\'e 1995). Like Kormendy
(1990) we notice an increase in $\rho _0$ 
 for lower luminosity galaxies.
But again, like in Figure 2, there is a large dispersion within the dIrrs. 
This seems to
indicate that galaxies with similar optical properties can have
dark matter halos with fairly different properties.
Athanassoula et al (1987) also suggested that halos of early-type
galaxies are more concentrated than those of later types. We also see a
a trend that the ratio of the core radius over $R_{25}$ increases for our 
dwarfs (but with a large spread here too). 

These trends have important implications for galaxy formation scenarios.
Indeed the CDM halos from N-body simulations with $\Omega _0 =1$ of Navarro,
Frenk \& White (1996)  are not compatible with observed rotation curves: the 
CDM halos are substantially more concentrated than what is inferred from
observations. Navarro, Eke \& Frenk (1996) have proposed that early
bursts of star formation could expell a large fraction of the baryonic
material in dwarfs therefore significantly altering the central
regions. But low-surface-brightness galaxies, which can have quite large
scalelengths and be massive objects (de Blok, this volume) are also
obsverved to be less concentrated than their simulated halos; since they
do not have the shallow potentials of dwarfs it is more difficult to
create baryonic outflows to solve their concentration problem. Navarro
(this volume) proposes instead that lower concentration halos, compatible
with dwarfs and LSBs observed curves, can be produced by low $\Omega _0$ flat  
CDM models, with $\Omega _0 =0.3$ and $\Lambda =0.7$;  
it should be noticed that the inclusion of a compensating $\Lambda$
term is mandatory for $\Omega < 1$ models, since the way current simulations
are constructed requires a flat geometry of the background cosmology
(Buchert 1995).

\begin{figure}[h]
\epsfxsize=8.5cm
\epsfysize=7cm
\centerline{\epsffile{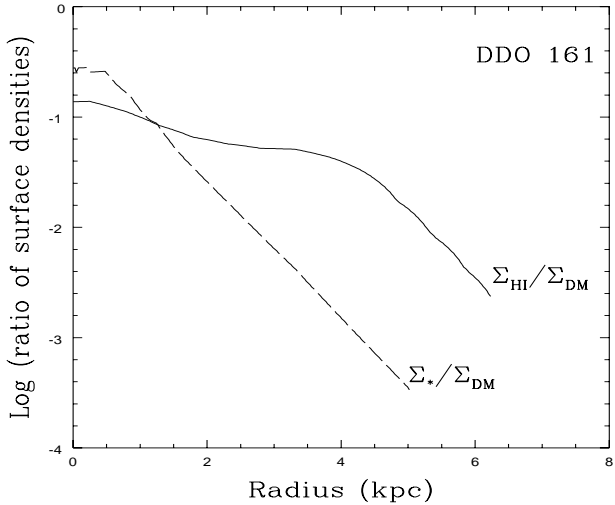}}
\caption[1]{\label{f4} Ratio of HI to dark matter
surface densities (full line), and stellar to 
dark matter densities (dashed line) for DDO 161}
\end{figure}

Another possible clue about the nature of dark matter comes from the
fact that in  most spiral galaxies the ratio of the HI 
surface densities to dark matter surface densities 
($\Sigma_{HI}/\Sigma_{DM}$) are seen to stay remarkably
constant (Bosma 1978), even out to large radii (Carignan et al 1990). 
This has been used as an argument for a strong coupling between the HI gas 
and the dark matter, 
 hinting that the dark matter is not dissipationless, therefore
 has possibly a baryonic nature.
One can then model the rotation curves using a scaled-up version of the
HI disks (ie: varying the gas mass) rather than a dark matter halo 
(see van Albada, this volume) and obtain reasonable fits (sometimes even
better ones, Broeils 1992).
In our dwarfs however this is no longer true: 
the ratio of HI to dark matter surface densities start dropping appreciably 
at roughly the Holmberg radius. Figure 4 shows the case of DDO 161 (for
which $R_{HO} \sim $3 kpc) where this ratio $\Sigma_{HI}/\Sigma_{DM}$ drops 
 by at least a factor
of 10 (similar factors are found for our other 4 dwarfs).  Many spirals
have HI radial profiles measured out to a larger number of $\alpha
^{-1}$ than some of our dwarfs (see the whole sample of Figure 3 for
example) and do not exhibit this decline in $\Sigma_{HI}/\Sigma_{DM}$. 
So this could imply that not only different galaxies can have different
fractions of dark matter but possibly a different mixture of dark matter
flavours (different fractions of baryonic and non-baryonic dark matter).

\section{Conclusions}

Dwarf irregular galaxies are dark-matter dominated, the dark matter
halos of our dwarfs account for 54\% up to 92\% of their total mass
(inside r$_{max}$).
In most cases the dark halo is the major dynamical contributor already
in the rising part of the rotation curve, and sometimes even the HI disk
is more massive than the stellar disk.

Our mass-models results show that these lower luminosity galaxies have
higher total mass-to-light ratios, and that their dark halos have higher
central densities and are less concentrated, confirming the Kormendy
(1989) correlations.

Contrary to what is found in normal spirals, the ratio of HI to dark matter 
surface densities are no longer constant
at large galactocentric radii. One can no longer fit scaled-up HI disks
instead of dark halos to explain the rotation curves, since at large
radii there is no longer a strong coupling between the HI gas and the
dark matter.

\acknowledgments

Thanks to ATNF and MSO TACs and Miller Goss for lots of
telescope time. Thanks to ANU and Fonds FCAR (Qu\'ebec) for 
financial support.

\end{document}